\documentclass[sn-mathphys-num]{sn-jnl}

\usepackage{graphicx}%
\usepackage{multirow}%
\usepackage{amsmath,amssymb,amsfonts}%
\usepackage{amsthm}%
\usepackage{mathrsfs}%
\usepackage[title]{appendix}%
\usepackage{xcolor}%
\usepackage{textcomp}%
\usepackage{manyfoot}%
\usepackage{mathtools}
\usepackage{booktabs}%
\usepackage{algorithm}%
\usepackage{algorithmicx}%
\usepackage{algpseudocode}%
\usepackage{listings}%

\theoremstyle{thmstyleone}%
\newtheorem{theorem}{Theorem}
\newtheorem{corollary}{Corollary}
\theoremstyle{thmstyletwo}%

\theoremstyle{thmstylethree}%

\begin{document}

\title[PT-Symmetric $SU(2)$ - like Random Matrix Ensembles: Invariant Distributions and Spectral Fluctuations]
{PT-Symmetric $SU(2)$ - like Random Matrix Ensembles: Invariant Distributions and Spectral Fluctuations}


\author[1,2]{\fnm{Stalin} \sur{Abraham}}\email{stalin.abraham@cbs.ac.in}
\equalcont{These authors contributed equally to this work.}

\author*[1,2]{\fnm{A.} \sur{Bhagwat}}\email{ameeya@cbs.ac.in}
\equalcont{These authors contributed equally to this work.}

\author[1]{\fnm{Sudhir Ranjan} \sur{Jain}}\email{srjain@cbs.ac.in}
\equalcont{These authors contributed equally to this work.}

\affil*[1]{\orgdiv{School of Physical Sciences}, \orgname{UM-DAE Centre for Excellence in Basic Sciences, 
University of Mumbai}, \orgaddress{\street{Vidyanagari Campus, Santacruz East}, \city{Mumbai}, \postcode{400098}, 
\state{Maharashtra}, \country{India}}}

\affil[2]{\orgdiv{Centre for Excellence in Theoretical and Computational Sciences}, 
\orgname{University of Mumbai}, \orgaddress{\street{Vidyanagari Campus, Santacruz East}, \city{Mumbai}, 
\postcode{400098}, \state{Maharashtra}, \country{India}}}


\abstract{
We consider an ensemble of $2\times 2$ normal matrices with complex entries representing 
operators in the quantum mechanics of 2 - level parity-time reversal (PT) symmetric
systems. 
The randomness of the ensemble is endowed by obtaining probability distributions based on 
symmetry and statistical independence. The probability densities turn 
out to be power law with exponents that depend on the boundedness of the domain. For small spacings, $\sigma$, 
the probability density varies as $\sigma^{\nu}$, $\nu \geq 2$. The degree of level repulsion is a 
parameter of great interest as it makes a connection to quantum chaos; the lower bound of $\nu$ for our 
ensemble coincides with the Gaussian Unitary Ensemble. We believe that the systematic development 
presented here paves the way for further generalizations in the field of random matrix theory for 
PT-symmetric quantum systems.}




\maketitle

\section{Introduction}\label{sec1}

Symmetries guide mathematical classification of physical systems \cite{wigner}. 
Among them, discrete symmetries like time reversal and parity play a special role. One of these appears in the statistical 
treatment of quantum systems with complex spectra, often mathematically treated by random matrix theory \cite{mehta,jg}. 
For systems where parity and time-reversal are broken ($PT$-symmetric), random matrix theory was introduced \cite{aj2003,aj2003_1}. 
These works introduced the concept of pseudo-unitary symmetry in the context of random matrices and proved the 
existence of a new universality class with novel level repulsion. The physical situations related to these random matrix 
models are diverse and most interesting. Historically, these models appeared in the context of quantum field theory 
with an indefinite metric \cite{sudarshan} and in the electromagnetic theory of transmission lines \cite{pease}. 
By introducing an indefinite metric, a general criterion for a set of non-Hermitian operators was established \cite{geyer}, 
consistent with the conventional interpretation of quantum mechanics \cite{jain2009}. The momentum distribution of 
an ideal gas of identical particles in two dimensions is an open problem. One of the ways to obtain this was shown by 
employing the eigenstate ensemble exploiting quantum chaos \cite{alonso_jain1,alonso_jain2} where the second virial 
coefficient for such gas was expressed in terms of a counting problem in the theory of Braid groups. Such systems 
are related on the one hand with Nambu's \cite{nambu} proposal of quark confinement in the medium of monopoles and 
with Aharonov-Bohm type quantum billiards which are pseudo-integrable \cite{djm}. The latter work paved the way 
towards a new universality class of random matrices for explaining the spectral fluctuations of pseudo-integrable and 
almost integrable quantum billiards \cite{gj}. Subsequently, these results have been generalized for 
$N \times N$ random cyclic matrices \cite{sri1,sri2}. Yet another fascinating connection is made with the exactly 
solvable models of many-particle systems. The joint probability distribution function of the eigenvalues of several 
random matrix ensembles coincides with the joint probability distribution function obtained from the many-body 
ground state \cite{srjain2006,rey}. A similar connection for random cyclic matrices exists \cite{sri3} where the 
connection was made with a screened harmonic oscillator. By a further transformation employing the normal 
coordinates, eigenfunctions of integrable \cite{rey} and chaotic billiards \cite{jgk} have been obtained.    

Here we re-visit the random matrix theory for non-Hermitian matrices. There has been a lot of work on random matrices 
with complex elements in the last sixty years \cite{ginibre,haake,forrester}. In all these works, the real and imaginary parts 
of the matrix elements were drawn from a Gaussian distribution. In the celebrated work pioneered by Wigner, Mehta, 
Gaudin, Dyson and others, this distribution was derived for the matrix ensembles based on time-reversal invariance, 
rotational symmetry, and statistical independence \cite{mehta}. However, for the case of non-Hermitian matrices, 
the distribution has been assumed to be Gaussian. As one of the main results, we present here a symmetry-based argument for deriving a probability distribution for the ensemble of non-Hermitian matrices considered here. 
Subsequently, we calculate the fluctuation properties of the eigenvalues and obtain new results. This leads to 
a degree of level repulsion, which is shown to be greater than or equal to two. It may be recalled that the degree of level repulsion is qualitatively associated with quantum chaos. The degree of level repulsion measures the rate 
at which the two eigenvalues go away from each other in the limit of small spacing. For the invariant Gaussian 
ensembles, the degree of repulsion is respectively 1, 2, and 4 for Orthogonal, Unitary, and Symplectic 
ensembles \cite{mehta}. It was found to be stronger for certain pseudo-Hermitian ensembles some twenty 
years ago \cite{aj2003,aj2003_1}. 

Specifically, in this work, we focus on the $2\times 2$ parity and time-reversal symmetric (PT-symmetric) complex matrices 
that are non-Hermitian but invertible and normal. It is worth pointing to a recent review \cite{bender_RMP} 
of the subject of PT-symmetric quantum mechanics, our effort falls very much into the physics and mathematics of the theme. 
These are interesting due to the fact that they can have real or complex eigenvalues and are used to understand, 
non - dissipative and dissipative systems, respectively 
(see, for example, \cite{BEN.98,BEN.99,BEN.03,AM.02a,AM.02b,SA.24}).

\section{Random matrix ensemble}\label{sec2}

We consider an ensemble of normal $2\times 2$ matrices with complex entries expressed as
\begin{eqnarray}\label{eq:H} 
H = 
\begin{pmatrix*}[r]
z_1 & z_2 \\
-\bar{z}_2 & \bar{z}_1 
\end{pmatrix*}
\end{eqnarray} 
with $z_1$ and $z_2$ are complex numbers, $z_1 = \left(x_1,y_1\right)$, and 
$z_2 = \left(x_2,y_2\right)$. These matrices are complex normal matrices and hence are 
unitarily diagonalisable. Further, they satisfy $H^{\dagger}H = HH^{\dagger} = \det(H) I_{2}$
where, $I_{2}$ is an identity matrix. Notice that $\det(H) = z_1\bar{z}_1 + z_2\bar{z_2}\ne 0 $,
unless $z_1 = z_2 = 0$. If $\det(H) = 1$, these would have been $SU(2)$ matrices. With this
motivation, the matrices considered in this analysis are called $SU(2)$-like matrices.

Suppose that $\epsilon \in \mathbb{R}$ is a very small number.
We demand invariance of probability distribution under similarity transformations.
Let us write an infinitesimal transformation,
\begin{eqnarray} 
Q = 
\begin{pmatrix*}[r]
1 & -\epsilon \\
\epsilon & 1 
\end{pmatrix*}
\end{eqnarray} 
under which, $H$ transforms as 
\begin{eqnarray} 
Q^{-1}HQ  = 
\begin{pmatrix*}[r]
z_1 + \epsilon\left(z_2 - \bar{z}_2\right) & z_2 + \epsilon\left(\bar{z}_1 - z_1\right)   \\
-z_2 + \epsilon\left(\bar{z}_1 - z_1\right) & \bar{z}_1 + \epsilon\left(\bar{z}_2 - z_2\right) 
\end{pmatrix*}
\end{eqnarray} 

\subsection{Construction of Probability Distribution Function}
Let $P(z)$ denote the probability distribution function of the parameters defining the matrix elements. The invariance of $P(z)$ under similarity
transformation implies that
\begin{eqnarray} 
P\left(z_1 + \epsilon\left(z_2 - \bar{z}_2\right)\right)
P\left(z_2 + \epsilon\left(\bar{z}_1 - z_1\right)\right) = P\left(z_1\right)P\left(z_2\right)
\end{eqnarray} 
We now make two assumptions: 1) $P(z) = f\left(x^2 + y^2\right) = f_1\left(x^2\right)f_2\left(y^2\right)$ 
with $z\in\mathbb{C}$ and $z=(x,y)$ and $f_1$, $f_2$ sufficiently smooth, and 
2) $P\left(z_1z_2\right) = P\left(z_1\right)P\left(z_2\right)$.
With these assumptions, we get:
\begin{eqnarray} 
P\left(z_1 + \epsilon\left(z_2 - \bar{z}_2\right)\right) = P\left(x_1 + iy_1 + i2\epsilon y_2\right) = 
f_1\left(x_{1}^{2} + y_{1}^{2} + 4\epsilon y_1 y_2\right) \\
P\left(z_2 + \epsilon\left(\bar{z}_1 - z_1\right)\right) = P\left(x_2 + iy_2 - i2\epsilon y_1\right) = 
f_2\left(x_{2}^{2} + y_{2}^{2} - 4\epsilon y_1 y_2\right)
\end{eqnarray} 
with 
\begin{eqnarray} 
f_1\left(x_{1}^{2} + y_{1}^{2} + 4\epsilon y_1 y_2\right) = f_1\left(x_{1}^{2} + y_{1}^{2}\right) + 
8\epsilon y_1 y_2 \left(x_1 \frac{\partial f_1}{\partial x_1} + y_1 \frac{\partial f_1}{\partial y_1}\right) + o\left(\epsilon^2\right) \\
f_2\left(x_{2}^{2} + y_{2}^{2} - 4\epsilon y_1 y_2\right) = f_2\left(x_{2}^{2} + y_{2}^{2}\right) - 
8\epsilon y_1 y_2 \left(x_2 \frac{\partial f_2}{\partial x_2} + y_2 \frac{\partial f_2}{\partial y_2}\right)  + o\left(\epsilon^2\right)
\end{eqnarray} 
Invariance of probability distribution function under similarity transformation requires that:
\begin{eqnarray} 
f_1\left(x_{1}^{2} + y_{1}^{2} + 4\epsilon y_1 y_2\right)
f_2\left(x_{2}^{2} + y_{2}^{2} - 4\epsilon y_1 y_2\right) = 
f_1\left(x_{1}^{2} + y_{1}^{2}\right)f_2\left(x_{2}^{2} + y_{2}^{2}\right)
\end{eqnarray} 
giving us
\begin{eqnarray} 
f_2\left(x_{2}^{2} + y_{2}^{2}\right) 
\left(x_1 \frac{\partial f_1}{\partial x_1} + y_1 \frac{\partial f_1}{\partial y_1}\right) = 
f_1\left(x_{1}^{2} + y_{1}^{2}\right) 
\left(x_2 \frac{\partial f_2}{\partial x_2} + y_2 \frac{\partial f_2}{\partial y_2}\right).
\label{e1}
\end{eqnarray} 
Let us separate $f_{1,2}$ in $x-$ and $y-$ parts:

\begin{eqnarray} 
f_1\left(x_{1}^{2} + y_{1}^{2}\right) = F_{11}\left(x_{1}^{2}\right)F_{12}\left(y_{1}^{2}\right)\\
f_2\left(x_{2}^{2} + y_{2}^{2}\right) = F_{21}\left(x_{2}^{2}\right)F_{22}\left(y_{2}^{2}\right)
\label{e2}
\end{eqnarray} 
Combining Eqs. (\ref{e1},\ref{e2}), we get:
\begin{eqnarray} 
F_{21}F_{22} \left[x_1 \frac{dF_{11}}{dx_1}F_{12} + y_1 \frac{dF_{12}}{dy_1}F_{11}\right] = 
F_{11}F_{12} \left[x_2 \frac{dF_{21}}{dx_2}F_{22} + y_2 \frac{dF_{22}}{dy_2}F_{21}\right]
\end{eqnarray} 
This leads to:
\begin{eqnarray} 
\frac{x_1}{F_{11}} \frac{dF_{11}}{dx_1} + \frac{y_1}{F_{12}} 
\frac{dF_{12}}{dy_1} = 
\frac{x_2}{F_{21}} \frac{dF_{21}}{dx_2} + \frac{y_2}{F_{22}} 
\frac{dF_{22}}{dy_2}
\end{eqnarray} 
Using the separation of variables, 
\begin{eqnarray} 
\frac{x_1}{F_{11}} \frac{dF_{11}}{dx_1} + \frac{y_1}{F_{12}} 
\frac{dF_{12}}{dy_1}
- \frac{x_2}{F_{21}} \frac{dF_{21}}{dx_2} = \frac{y_2}{F_{22}} \frac{dF_{22}}{dy_2} = \lambda_0
\in \mathbb{R}
\end{eqnarray} 
where $\lambda_0$ is a separation constant. 
This leads directly to
\begin{eqnarray} 
F_{22}\left(y_2\right) = y_{2}^{\lambda_0}
\end{eqnarray} 
Further,
\begin{eqnarray} 
\frac{x_1}{F_{11}} \frac{dF_{11}}{dx_1} + \frac{y_1}{F_{12}} 
\frac{dF_{12}}{dy_1}
= \lambda_0 + \frac{x_2}{F_{21}} \frac{dF_{21}}{dx_2} = \lambda_2 \in \mathbb{R}
\end{eqnarray} 
leading to:
\begin{eqnarray} 
F_{21}\left(x_2\right) = x_{2}^{\lambda_2 - \lambda_0}.
\end{eqnarray} 
Also,
\begin{eqnarray} 
\frac{x_1}{F_{11}} \frac{dF_{11}}{dx_1} = \lambda_2 - \frac{y_1}{F_{12}} \frac{dF_{12}}{dy_1}
 = \lambda_3 \in \mathbb{R}.
\end{eqnarray} 
Finally, 
\begin{eqnarray} 
F_{11}\left(x_1\right) = x_{1}^{\lambda_3} \quad \mathrm{and} \quad
F_{12}\left(y_1\right) = y_{1}^{\lambda_2 - \lambda_3}
\end{eqnarray} 
In summary, we get the probability distribution function:
\begin{eqnarray} 
P\left(z_1,z_2\right) = P\left(x_1,y_1,x_2,y_2\right) = Ax_{1}^{\lambda_3}y_{1}^{\lambda_2 - \lambda_3} 
x_{2}^{\lambda_2 - \lambda_0}y_{2}^{\lambda_0}
\end{eqnarray} 

\subsection{Determination of Normalisation Constant}
Let $\left(x_1,y_1,x_2,y_2\right) \in \mathbb{R}^{4}$. Take the weight function
$w : \mathbb{C}\times\mathbb{C} \rightarrow \mathbb{R}$ which is given by
\begin{eqnarray} 
w\left(z_1,z_2\right) = w\left(x_1,y_1,x_2,y_2\right) &=& \exp\left\{-\alpha \left(\left|z_1\right|^2 + 
\left|z_2\right|^2\right)\right\} \\ 
 &=& \exp\left\{-\alpha \left(x_{1}^2 + y_{1}^{2} + x_{2}^{2} + y_{2}^{2} \right) \right\}
\end{eqnarray} 
with $\alpha > 0$. We demand that
\begin{eqnarray} 
\int_{\mathbb{R}^4} P\left(x_1,y_1,x_2,y_2\right) w\left(x_1,y_1,x_2,y_2\right) d^{4}x = 1
\end{eqnarray} 
where $d^{4}x :=dx_1dx_2dy_1dy_2$.
Explicitly,
\begin{eqnarray} 
A \int_{\mathbb{R}} x_{1}^{\lambda_3}e^{-\alpha x_{1}^{2}} dx_1 
  \int_{\mathbb{R}} y_{1}^{\lambda_2 - \lambda_3}e^{-\alpha y_{1}^{2}} dy_1 
  \int_{\mathbb{R}} x_{2}^{\lambda_2 - \lambda_0}e^{-\alpha x_{2}^{2}} dx_2 
  \int_{\mathbb{R}} y_{2}^{\lambda_0}e^{-\alpha y_{2}^{2}} dy_2 = 1
\end{eqnarray} 
It is well known that
\begin{eqnarray} 
\int_{\mathbb{R}} x^{a} e^{-bx^2} \,dx = \frac{1}{2}\left[1 + \left(-1\right)^a\right] b^{-\frac{a+1}{2}} \,
\Gamma\left(\frac{a+1}{2}\right) ~ \mathrm{with} ~ \Re a > -1 ~ \mathrm{and} ~ \Re b > 0
\end{eqnarray}
where $\Gamma$ is the usual gamma function. Now, by our assumption, $ a, b\in\mathbb{R}$. Thus, the 
above integrals are convergent when $a > -1 $ and $b > 0$ in all the cases. 
Explicitly, we have:
\begin{eqnarray} 
\int_{\mathbb{R}} x_{1}^{\lambda_3} \,e^{-\alpha x_{1}^{2}} dx_1 &=& 
\left[1 + \left(-1\right)^{\lambda_3}\right]
\alpha^{-\left(\lambda_3 + 1\right)/2}\,\,\Gamma\left(\frac{\lambda_3 + 1}{2}\right) \\
\int_{\mathbb{R}} x_{2}^{\lambda_2 - \lambda_0} \,e^{-\alpha x_{2}^{2}} dx_2 &=& 
\left[1 + \left(-1\right)^{\lambda_2 - \lambda_0}\right]
\alpha^{-\left(\lambda_2-\lambda_0+1\right)/2}
 \,\,\Gamma\left(\frac{\lambda_2 - \lambda_0 + 1}{2}\right) \\
\int_{\mathbb{R}} y_{1}^{\lambda_2 - \lambda_3} \,e^{-\alpha y_{1}^{2}} dy_1 &=& 
\left[1 + \left(-1\right)^{\lambda_2 - \lambda_3}\right] 
\alpha^{-\left(\lambda_2- \lambda_3+1\right)/2}
\,\,\Gamma\left(\frac{\lambda_3 - \lambda_2 + 1}{2}\right) \\
\int_{\mathbb{R}} y_{2}^{\lambda_0} \,e^{-\alpha y_{2}^{2}} dy_2 &=& 
\left[1 + \left(-1\right)^{\lambda_0}\right] 
\alpha^{\left(\lambda_0+1\right)/2} \,\,
\Gamma\left(\frac{\lambda_0 + 1}{2}\right)
\end{eqnarray} 
In order that the above integrals are convergent, we need to demand that $\lambda_3 > -1$, 
$\lambda_2 - \lambda_0 > -1$, $\lambda_2 - \lambda_3 > -1$ and $\lambda_0 > -1$. Further, in order
that the integrals are non-zero, we need
\begin{eqnarray} 
\left(-1\right)^{\lambda_3} \ne -1 \Rightarrow \lambda_3 \ne 2n + 1 \quad \forall n \in \mathbb{Z} \\
\left(-1\right)^{\lambda_2 - \lambda_0} \ne -1 \Rightarrow \lambda_2 - \lambda_0 \ne 2n + 1 \quad \forall n \in \mathbb{Z} \\
\left(-1\right)^{\lambda_2 - \lambda_3} \ne -1 \Rightarrow \lambda_2 - \lambda_3 \ne 2n + 1 \quad \forall n \in \mathbb{Z} \\
\left(-1\right)^{\lambda_0} \ne -1 \Rightarrow \lambda_0 \ne 2n + 1 \quad \forall n \in \mathbb{Z}
\end{eqnarray} 
Thus, in conclusion, the above integrals exist if $\lambda_0 > -1$, $\lambda_2 > -2$ and $\lambda_3 > -1$ with an additional constraint that none of these should be odd integers. Notice further that 
$(-1)^{\beta} \in \mathbb{C}$ unless $\beta \in \mathbb{Z}$. Thus, $\lambda_3$ and $\lambda_0$ are 
integers. However, these cannot be odd integers, hence $\lambda_0$ and $\lambda_3$ are even integers. 
Let $\lambda_0 = 2n$ and $\lambda_3 = 2l$ such that $n,l \in \mathbb{N}\cup\{0\}$. For the sake of brevity, we
will denote $\mathbb{N}_{0} = \mathbb{N}\cup\{0\}$. Along the same lines, it follows that $\lambda_2 - \lambda_0$
should be an integer, giving us, $\lambda_2$ is an integer, in fact, an even integer. Further, given that
$\lambda_2 - \lambda_0 > -1$, if $\lambda_2 = 2m$, $m\in\mathbb{Z}$ then $2m - 2n > -1$ $\Rightarrow$ 
$m \ge n$. Similarly, $\lambda_2 - \lambda_3 > -1$ gives $m\ge l$. Thus, it follows that 
$m \ge \max\{l,n\}$. In conclusion, $\lambda_0 = 2n$, $\lambda_2 = 2l$ and $\lambda_3 = 2m$ with 
$l,m,n \in \mathbb{N}_{0}$ and $m \ge \max\{l,n\}$.  
  
With all these simplifications, and by using properties of Gamma function \cite{roy}, we get:
\begin{eqnarray} 
A_{lnm} = \frac{2^{2m}\alpha^{2m + 2}}{\pi^2 \xi_{lnm}}
\end{eqnarray} 
with 
\begin{eqnarray} 
\xi_{lnm} = (2l - 1)!!\,(2n - 1)!!\,(2m - 2n - 1)!!\,(2m - 2l - 1)!!
\end{eqnarray} 
where the symbols have their usual meanings. This finally yields,
\begin{eqnarray} 
P\left(x_1,y_1,x_2,y_2\right) = A_{lnm}x_{1}^{2l}y_{1}^{2(m-l)}x_{2}^{2(m-n)}y_{2}^{2n}
\end{eqnarray} 
such that $l,m,n \in \mathbb{N}_{0}$ and $m \ge \max\{l,n\}$.

\subsection{Level Repulsion}

The matrix $H$ (Eq. \eqref{eq:H}) has two eigenvalues, 
\begin{eqnarray} 
E_{\pm} = x_{1} \pm i \sqrt{y_{1}^{2} + y_{2}^{2} + x_{2}^2} \, .
\end{eqnarray}  
The level repulsion can be computed using the joint probability distribution:
\begin{eqnarray} 
\mathcal{P}(\sigma) = \int_{\mathbb{R}^{4}} \delta\left(\left|E_{+} - E_{-}\right| - \sigma\right) 
P\left(x_1,y_1,x_2,y_2\right) w\left(x_1,y_1,x_2,y_2\right)d^{4}x
\end{eqnarray} 
where the function $w$ is the weight function defined above and $\delta$ is the usual Dirac delta measure 
(see, for example, \cite{GEL}). 
Explicitly,
\begin{eqnarray} 
\mathcal{P}(\sigma) = \frac{1}{2} A_{lnm}\int_{\mathbb{R}^{4}} 
\delta\left(\sqrt{y_{1}^{2} + y_{2}^{2} + x_{2}^2} - \sigma/2\right) 
x_{1}^{2l}y_{1}^{2(m-l)}x_{2}^{2(m-n)}y_{2}^{2n} \nonumber \\ 
\times ~ e^{-\alpha\left(y_{1}^{2} + y_{2}^{2} + x_{2}^2 + x_{1}^{2}\right)}d^{4}x
\end{eqnarray} 
A close inspection of the above integral reveals that transformation to spherical polar coordinates could be 
useful here. Specifically, let $R^2 = y_{1}^{2} + y_{2}^{2} + x_{2}^2$ such that 
$y_1 = R \sin\theta\cos\phi$, $y_2 = R \sin\theta\sin\phi$, and $x_2 = R\cos\theta$. Using this transformation,
the above integral becomes 
\begin{eqnarray} 
\mathcal{P}(\sigma) &=& \frac{1}{2}A_{lnm}\int_{\mathbb{R}}x_{1}^{2l}e^{-\alpha x_{1}^{2}} dx_1
\int_{0}^{\infty}R^{4m - 2l + 2} e^{-\alpha R^2} \delta\left(R - \sigma/2\right)\,dR \nonumber \\
&& \times \int_{0}^{2\pi} \left(\cos\phi\right)^{2m - 2l} \left(\sin\phi\right)^{2n}\,d\phi 
\int_{0}^{\pi} \left(\sin\theta\right)^{2m + 2n - 2l + 1}\left(\cos\theta\right)^{2m - 2n}d\theta
\end{eqnarray} 
This integral can be evaluated analytically using the properties of beta integral, and the 
integral representation of the Gamma function (see, for example, \cite{roy}), leading to
\begin{eqnarray} 
\mathcal{P}(\sigma) = \frac{1}{2} \frac{\alpha^{2m - l + 3/2}}{\sqrt{\pi}\left(2m - l + 1/2\right)}
\left(\frac{\sigma^{2}}{2}\right)^{2m - l + 1}e^{-\alpha\sigma^2/4}
\end{eqnarray} 
which is the desired result. The degree of level repulsion is given by the exponent on $\sigma$ as $\sigma \to 0$, in this case it is $\sigma ^{2(2m - l + 1)}$. Recalling the admissible values of $m, l$, we see that the minimum value of $m$ is $l$; this gives the asymptotic behaviour as $\sigma^{2(l+1)}$. Thus, for the minimum value of $l = 0$, the degree of level repulsion coincides with that of the Gaussian Unitary Ensemble \cite{mehta}. For any other value, the degree of level repulsion is stronger than 2. This is an algebraic power, in contrast to the case of pseudo-Hermitian matrices where, for one case, it was shown to vary as $\sigma \log (1/\sigma)$ \cite{aj2003}.  

\subsection{Probability Distribution Function and Level Repulsion in a bounded Region of $\mathbb{R}^4$}

We begin with the expression for the probability distribution (up to the normalisation factor) obtained above:
\begin{eqnarray} 
P\left(x_1,y_1,x_2,y_2\right) = Ax_{1}^{\lambda_3}y_{1}^{\lambda_2 - \lambda_3} 
x_{2}^{\lambda_2 - \lambda_0}y_{2}^{\lambda_0}
\end{eqnarray} 
For the sake of convenience, we re-label some variables here. Out of the four variables, we will re-label 
$y_1$ to $x_3$ and $y_2$ to $x_4$. This leads to:
\begin{eqnarray} 
P\left(x_1,x_2,x_3,x_4\right) = Ax_{1}^{\lambda_3}x_{3}^{\lambda_2 - \lambda_3} 
x_{2}^{\lambda_2 - \lambda_0}x_{4}^{\lambda_0}
\end{eqnarray} 
Let $V\subset\mathbb{R}^4$ be defined by $x_i \ge 0$ for $i = 1, 2, 3, 4$, such that 
\begin{eqnarray} 
\sum_{i=1}^{4}x_{i}^{2} \le 1
\end{eqnarray} 
We first state a result due to Liouville that essentially generalises the beta integral to dimensions 
higher than 2 (see \cite{roy} for details):
\begin{theorem}
If $V$ is a region defined by $x_i \ge 0$, $i = 1, 2, ..., n$ and $\sum x_i \le 1$, then for 
$\Re \alpha_i > 0$, 
\begin{eqnarray} 
\int_{V} \prod_{i=1}^{n} x_{i}^{\alpha_{i}-1} dx_1...dx_n = \frac{\prod_{i=1}^{n}\Gamma\left(\alpha_i\right)}
{\Gamma\left(1 + \sum_{i=1}^{n}\alpha_{i}\right)}
\end{eqnarray} 
\end{theorem}
As an immediate corollary to this theorem, we have \cite{roy}:
\begin{corollary}
If $V$ is the region enclosed by $x_i \ge 0$ and $\sum \left(x_i/a_i\right)^{p_{i}} \le 1$ then
for $\Re\alpha_i > 0$,
\begin{eqnarray} 
\int_{V} \prod_{i=1}^{n} x_{i}^{\alpha_{i}-1} dx_1...dx_n = \frac{\prod_{i=1}^{n}\left(a_{i}^{\alpha_{i}}/p_{i}\right)
\Gamma\left(\alpha_i/p_i\right)}{\Gamma\left(1 + \sum_{i=1}^{n}\alpha_{i}/p_i\right)}
\end{eqnarray} 
\end{corollary}
The normalisation constant in the present case can be obtained readily from this corollary by 
setting $p_i = 2 $ and $a_i = 1$. Explicitly,
\begin{eqnarray} 
A\int_{V} \prod_{i=1}^{4} x_{i}^{\alpha_{i}-1} d^{4}x = \frac{1}{16}\frac{\prod_{i=1}^{n}
\Gamma\left(\alpha_i/2\right)}{\Gamma\left(1 + \sum_{i=1}^{n}\alpha_{i}/2\right)}
\end{eqnarray} 
giving us,
\begin{eqnarray} 
A = \frac{16\Gamma\left(1 + \sum_{i=1}^{n}\alpha_{i}/2\right)}{\prod_{i=1}^{n}\Gamma\left(\alpha_i/2\right)}
\end{eqnarray} 
where, $\alpha_1 = \lambda_3 + 1$, $\alpha_2 = \lambda_2 - \lambda_0 + 1$, $\alpha_3 = \lambda_2 - \lambda_3 + 1$ and 
$\alpha_4 = \lambda_0 + 1$ and $d^{4}x := dx_1dx_2dx_3dx_4$. 
Given the condition that $\Re\alpha_i>0$, in our case, the above integral exists if and 
only if $\lambda_3 + 1 > 0$, $\lambda_2 - \lambda_3 + 1 > 0$, $\lambda_2 - \lambda_0 + 1 > 0$, and $\lambda_0 + 1 > 0$.
Thus, it follows that the integral and hence the normalisation constant exists provided that $\lambda_3 > -1$, $\lambda_0 > -1$ and 
$\lambda_2 > \max\left\{\lambda_0,\lambda_3\right\}$. Further notice that for the domain over which $\lambda_k$ are defined,
the normalisation constant is always non-zero, as it should be. There is no further restriction on the values that 
the separation constants can take, which is in 
contrast with the case of full $\mathbb{R}^{4}$, where the separation constants were restricted to positive even integral values.

We now compute the nearest-neighbour level spacing distribution in this case.  Specifically, we need to evaluate the following integral:
\begin{eqnarray} 
{\mathcal{P}}\left(\sigma\right) = \frac{1}{2} A \int_{V} 
\prod_{i=1}^{4} x_{i}^{\alpha_{i}-1}\delta\left(\sqrt{x_{2}^{2} + x_{3}^{2} + x_{4}^{2}} - \sigma/2\right)d^{4}x
\end{eqnarray} 
It is convenient to use hyper-spherical coordinates here \cite{blumenson}. Specifically, we set
\begin{eqnarray} 
x_1 &=& r \cos\phi_1 \\
x_2 &=& r \cos\phi_2\sin\phi_1 \\
x_3 &=& r \sin\theta\sin\phi_1\sin\phi_2 \\
x_4 &=& r \cos\theta\sin\phi_1\sin\phi_2
\end{eqnarray} 
such that $r^2 = \sum_{i=1}^{4} x_{i}^{2}$ ($0\le r \le 1$), $\phi_1 \in \left[0,\pi\right]$, $\phi_2\in\left[0,\pi\right]$, and
$\theta\in\left[0,2\pi\right]$. In the present case, since we are restricting ourselves to non-negative values of 
$x_i$, it follows that $\phi_1,\phi_2,\theta\in\left[0,\pi/2\right]$. Further, the determinant of the Jacobian matrix is
$r^3\left(\sin\phi_1\right)^{2}\sin\phi_2$, and $x_{2}^{2}+x_{3}^{2}+x_{4}^{2} = r^2\left(\sin\phi_1\right)^{2}$,
which leads to
\begin{eqnarray} 
{\mathcal{P}}\left(\sigma\right) &=& \frac{1}{2} A\int_{0}^{1}\int_{\xi}^{\pi/2}\int_{0}^{\pi/2}\int_{0}^{\pi/2}
\delta\left(r\sin\phi_1 - \sigma/2\right)r^{\sum_{i}\alpha_i - 1} 
\left(\sin\theta\right)^{\alpha_3 - 1}\left(\cos\theta\right)^{\alpha_4 - 1} \times
\nonumber \\
&& \hskip -5pt \left(\cos\phi_{2}\right)^{\alpha_2 - 1}\left(\sin\phi_{2}\right)^{\alpha_3 + \alpha_4 - 1}
\left(\cos\phi_1\right)^{\alpha_1 - 1}
\left(\sin\phi_{1}\right)^{\alpha_2 + \alpha_3 + \alpha_4 - 1} 
 drd\phi_1d\phi_2d\theta
\end{eqnarray} 
Notice that the domain of integral over $\phi_1$ has been cut to $\left[\xi,\pi/2\right]$ with $\xi > 0$. This is 
due to the constraint implied by the Dirac Delta measure, as will be explained below. 

We will first carry out integral over $r$. In order to do that, we first write:
\begin{eqnarray} 
{\mathcal{P}}\left(\sigma\right) &=& \frac{1}{2} A\int_{0}^{1}\int_{\xi}^{\pi/2}\int_{0}^{\pi/2}\int_{0}^{\pi/2}
\delta\left(r - \sigma/\left(2\sin\phi_1\right)\right)r^{\sum_{i}\alpha_i - 1} 
\left(\sin\theta\right)^{\alpha_3 - 1}\left(\cos\theta\right)^{\alpha_4 - 1} 
\nonumber \\
&\times& \hskip -5pt \left(\cos\phi_{2}\right)^{\alpha_2 - 1}\left(\sin\phi_{2}\right)^{\alpha_3 + \alpha_4 - 1}
\left(\cos\phi_1\right)^{\alpha_1 - 1}
\left(\sin\phi_{1}\right)^{\alpha_2 + \alpha_3 + \alpha_4 - 2} 
\,drd\phi_1d\phi_2d\theta
\end{eqnarray} 
leading to:
\clearpage
\begin{eqnarray} 
{\mathcal{P}}\left(\sigma\right) &=& \frac{1}{2}A\left(\frac{\sigma}{2}\right)^{\sum_{i}\alpha_i - 1} 
\int_{\xi}^{\pi/2}\left(\cos\phi_1\right)^{\alpha_1 - 1}\left(\sin\phi_{1}\right)^{-\alpha_1 - 1}d\phi_1 \nonumber \\
&\times& \int_{0}^{\pi/2}\left(\sin\theta\right)^{\alpha_3 - 1}\left(\cos\theta\right)^{\alpha_4 - 1}d\theta
\int_{0}^{\pi/2}\left(\cos\phi_{2}\right)^{\alpha_2 - 1}\left(\sin\phi_{2}\right)^{\alpha_3 + \alpha_4 - 1}d\phi_2 \nonumber \\
\end{eqnarray} 
Out of these, the second and the third are the standard beta integrals \cite{roy}, and are given by
\begin{eqnarray} 
\int_{0}^{\pi/2}\left(\sin\theta\right)^{\alpha_3 - 1}\left(\cos\theta\right)^{\alpha_4 - 1}d\theta =
\frac{\Gamma\left(\alpha_3/2\right)\Gamma\left(\alpha_4/2\right)}{2\Gamma\left(\left(\alpha_3 + \alpha_4\right)/2\right)}
\end{eqnarray} 
and
\begin{eqnarray} 
\int_{0}^{\pi/2}\left(\cos\phi_{2}\right)^{\alpha_2 - 1}\left(\sin\phi_{2}\right)^{\alpha_3 + \alpha_4 - 1}d\phi_2 = 
\frac{\Gamma\left(\left(\alpha_3+\alpha_4\right)/2\right)\Gamma\left(\alpha_2/2\right)}{2\Gamma\left(\left(\alpha_2+\alpha_3 + \alpha_4\right)/2\right)}
\end{eqnarray} 
In order to evaluate the first integral, we need to estimate $\xi$ first. We know that 
\begin{eqnarray} 
0 \le \frac{\sigma^{2}}{4\left(\sin\phi_1\right)^2} \le 1
\end{eqnarray} 
The lower limit of the integral, $\xi$, is obtained by demanding that it is the one that gives the smallest 
possible value of $\phi_1$, leading to $\sin\xi = \sigma/2$, giving us $\xi = \arcsin\left(\sigma/2\right)$. The first 
integral can now be evaluated:
\begin{eqnarray} 
\int_{\xi}^{\pi/2}\left(\cos\phi_1\right)^{\alpha_1 - 1}\left(\sin\phi_{1}\right)^{-\alpha_1 - 1}d\phi_1 = 
\frac{1}{\alpha_1}\left(\cot\xi\right)^{2} = \frac{4 - \sigma^2}{\sigma^2}
\end{eqnarray} 
Upon combining all these factors and inserting the value of normalisation constant, we finally get
\begin{eqnarray} 
{\mathcal{P}}\left(\sigma\right) = \frac{4\Gamma\left(1 + \sum_{i=1}^{4}\alpha_i/2\right)}
{\Gamma\left(\left(\alpha_2 + \alpha_3 + \alpha_4\right)/2\right)}\,\frac{4 - \sigma^2}{\alpha_1 \sigma^2}\,
\left(\frac{\sigma}{2}\right)^{\sum_{i}^{4}\alpha_i - 1}
\end{eqnarray} 
Finally, note that $\sum_i \alpha_i = 2\lambda_2 + 4$. Thus,
\begin{eqnarray} 
{\mathcal{P}}\left(\sigma\right) = \frac{4\Gamma\left(\lambda_2 + 3\right)}
{\Gamma\left(\lambda_2 - \lambda_3/2 + 3/2\right)}\,\frac{4 - \sigma^2}{\left(\lambda_3 + 1\right)}\,
\left(\frac{\sigma}{2}\right)^{2\lambda_2 + 1}
\end{eqnarray} 

\section{Concluding remarks}

As emphasized in the main text, one of the significant developments reported here is to obtain the 
distribution function of the elements of an ensemble of non-Hermitian $SU(2)$-like matrices on the basis of symmetry and 
statistical independence. Hitherto, it was always assumed to be a Gaussian distribution. For at least 
the ensemble considered here, this assumption has been lifted. This has led us to the consideration of 
certain cases dictated by the domain of the matrix elements. Employing the new-found probability distribution, 
we have calculated the nearest-neighbour level spacing distribution. The degree of level repulsion is 
tunable as there appear two parameters; this result is potentially significant in modelling physical 
situations where interaction strength varies (see, e.g. \cite{djm}) and also in connection with number 
theory (see \cite{riemann}). However, at the lower bound of the level repulsion parameter of our ensemble is the degree for the Gaussian Unitary ensembles (GUE). It is worth recalling that GUE corresponds to 
physical systems which violate time-reversal invariance. In line with earlier results 
\cite{aj2003,aj2003_1,sri1,sri2}, we are led to conjecture that an additional breakdown of parity 
makes the degree of level repulsion stronger; thus, we conjecture:
\begin{equation*}
\lim_{\sigma \to 0} P(\sigma) \sim \sigma ^{\nu}, \nu \geq 2.   
\end{equation*}

The connections between random matrix theory and exactly solvable models are profound 
\cite{srjain2006,gjk,c,b,jgk}. Upon appropriate generalization, we believe that this work 
opens up the possibility of connections with new, exactly solvable models also.

\backmatter

\bmhead{Acknowledgments}

SA acknowledges financial support through the Cyrus Guzder Fellowship. Valuable comments from the Doctoral Advisory Committee of UM-DAE Centre for Excellence in Basic Sciences are gratefully acknowledged.

\section*{Declarations}

\begin{itemize}
\item Funding: Partial financial support to Stalin Abraham was received through Cyrus Guzder Fellowship.
\item Conflict of interest/Competing interests: The authors have no competing interests to declare that are 
relevant to the content of this article.
\item Ethics approval and consent to participate: Not applicable.
\item Consent for publication: The authors give the publisher the due consent for publication of this article.
\item Data availability: Not applicable.
\item Materials availability: Not applicable.
\item Code availability: Not applicable.
\item Author contribution: The authors declare equal contribution to this article.
\end{itemize}



\end{document}